\documentclass[aps,twocolumn,prd]{revtex4-1}

\usepackage{bm,graphicx,textcomp,amssymb,amsmath,dcolumn,color}

\newcommand{\ket}[1]{\left|#1\right>}
\newcommand{\bra}[1]{\left<#1\right|}

\newcommand{\nn}{\nonumber\\}

\newcommand{\f}[1]{\mbox{\boldmath$#1$}}

\newcommand{\na}{\mbox{\boldmath$\nabla$}}
\newcommand{\bea}{\begin{eqnarray}}
\newcommand{\ea}{\end{eqnarray}}
\newcommand{\eea}{\end{eqnarray}}
\newcommand{\ord}{\,{\cal O}}

\begin{document}

\title{Energy transfer between gravitational waves and quantum matter}



\author{Jonathan Gr\"afe}

\author{Falk Adamietz}

\author{Ralf Sch\"utzhold}

\affiliation{Helmholtz-Zentrum Dresden-Rossendorf, 
Bautzner Landstra{\ss}e 400, 01328 Dresden, Germany,}

\affiliation{Institut f\"ur Theoretische Physik, 
Technische Universit\"at Dresden, 01062 Dresden, Germany,}

\date{\today}

\begin{abstract}
We study the interaction between gravitational waves and 
quantum matter such as Bose-Einstein condensates, super-fluid Helium, 
or ultra-cold solids, explicitly taking into account the changes of the 
trapping potential induced by the gravitational wave. 
As a possible observable, we consider the change of energy due to the 
gravitational wave, for which we derive rigorous bounds in terms of 
kinetic energy and particle number. 
Finally, we discuss implications for possible experimental tests. 
\end{abstract}

\maketitle

\section{Introduction} 

Gravitational waves had been predicted shortly after the publication of 
Einstein's field equations of general relativity \cite{Einstein-1916,Einstein-1918}, 
but their experimental detection was thought to be impossible for a long 
time in view of the smallness of the expected signals. 
It took roughly half a century until Weber constructed a detector for 
gravitational waves based on resonant mass antennas known as Weber bars \cite{Weber-1967,Weber-1968,Weber-1969}. 
Although Weber's initial results and claims of having detected a signal 
could not be reproduced by other groups, his endeavors should still be 
considered pioneering experiments, paving the way for later developments. 

A whole century after their prediction, gravitational waves have been detected 
at LIGO \cite{LIGO-paper,LIGO-www}, marking a major breakthrough 
and the beginning of a new era
in modern physics. 
Note that one should distinguish two major detection schemes for 
gravitational waves:
At interferometers such as LIGO, one measures the changes of the arm lengths 
and the resulting interference patters {\em during} the passage of the 
gravitational wave. 
In contrast, the resonant excitation of a Weber bar can be measured {\em after} 
the gravitational wave passed by. 

In the following, we theoretically investigate detection schemes of the 
second type, see also \cite{Berlin1,Berlin2,Berlin3}. 
Instead of Weber bars, we consider more general resonant mass antennas 
represented by quantum matter such as Bose-Einstein condensates, 
super-fluid Helium, or ultra-cold solids. 
%
To some extent, these studies are motivated by recent and partly controversial 
discussions regarding the use of Bose-Einstein condensates as gravitational 
wave detectors, see, e.g., \cite{Fagnocchi,Sabin,Ratzel:2018srb,Schutzhold-BEC,Howl-Comment,Schutzhold-Reply,Robbins-2019,Bruschi,Howl,Robbins-2022}.
More generally, the weakness 
of the interaction with gravitational waves 
and the resulting smallness of the signal 
motivates a quantum description.
The aforementioned examples for quantum
matter   
may offer certain advantages,   
e.g., regarding temperature, purity, or experimental control, 
see also \cite{Unruh-nondemolition,Unruh-linear,Detection,Helium,Ghayour,Vadakkumbatt,Shinn,Spengler:2021rlg}.

As a typical observable, we consider the change in energy 
induced by the gravitational wave. 
Note that matter-wave interferometers, which have also been proposed as  
gravitational wave detectors \cite{Stodolsky,Chiao,Speliotopoulos,Roura,Foffa,Gao-2011,Gao-2018}, are typically based on detection 
schemes of the first type -- and are thus not considered here.  

\section{Gravitational Waves} 

For simplicity, we consider linearly polarized gravitational waves propagating 
in a fixed direction. 
Other waves can be written as linear combinations of such solutions.
In a suitable coordinate system, the metric reads 
(using natural units $\hbar=c=\varepsilon_0=\mu_0=1$)
\bea
\label{metric}
\mathrm{d}s^2=\mathrm{d}t^2-[1+h]\,\mathrm{d}x^2-[1-h]\,\mathrm{d}y^2-\mathrm{d}z^2
\,,
\ea
where the function $h(t-z)$ describes the gravitational wave. 
However, as its wavelength is much larger than the characteristic length 
scales in the laboratory while its period is shorter than the duration of the 
experiment, we use the approximation $h(t-z)\approx h(t)$ in what follows.
Furthermore, since $h$ is extremely small, $h=\ord(10^{-22})$, we neglect 
quadratic terms $\ord(h^2)$ in the following 
(as usual in the linearized theory of gravitational waves).
As a consequence, the metric determinant can be approximated by unity 
$\sqrt{-g}=1+\ord(h^2)$. 

\subsection{Massive Particles}\label{Massive Particles}

Before investigating the implications of the metric~\eqref{metric} 
for the quantum Hamiltonian 
in the next Section~\ref{Matter Hamiltonian}, 
let us briefly discuss the impact on classical 
point particles and electromagnetic waves, which will also 
be relevant for changes of the trapping potential. 

Since the Christoffel symbols corresponding to Newton's gravitational 
acceleration vanish $\Gamma^i_{00}=0$, massive particles at rest with respect 
to the coordinates~\eqref{metric}, i.e., at constant positions $x$, $y$, 
and $z$, are solutions to the geodesic equations. 
As a result, the heavy mirrors used in LIGO, for example, do not change their 
positions $x$, $y$, and $z$ during the passage of the gravitational wave. 
However, their physical distance~\eqref{metric} changes, which can be 
measured by light rays, for instance.  

For moving particles, on the other hand, the gravitational wave does 
generate an effective force. 
Considering a non-relativistic motion in the $x,y$-plane for simplicity,
the Christoffel symbols $\Gamma^i_{0i}$ are given by $\pm\dot h/2$ and 
correspond to the acceleration $\dot u^x=-\dot h u^x$ and $\dot u^y=\dot h u^y$
in terms of the four-velocity $u^\mu$. 
The Christoffel symbols $\Gamma^0_{ij}$ then yield the change of energy 
$\dot u^t=\dot h(u_y^2-u_x^2)/2$.

\subsection{Electromagnetic Waves}\label{Electromagnetic Waves} 

Next, let us consider electromagnetic waves propagating in the background 
metric~\eqref{metric} which are described by the Maxwell equations 
$\nabla_\mu F^{\mu\nu}=0$ with the electromagnetic field-strength tensor 
$F_{\mu\nu}=\partial_\mu A_\nu-\partial_\nu A_\mu$ and the vector potential 
$A_\mu$. 
Their dispersion relation can already be read off the metric~\eqref{metric}
\bea
\label{dispersion}
\Omega^2=-g^{ij} K_i K_j=[1-h]K_x^2+[1+h]K_y^2+K_z^2
\,.
\ea
Since the changes of the amplitudes $A_i$ induced by the gravitational wave 
depend on their polarization, let us first consider the cases of fixed 
polarizations along the coordinate axes for simplicity. 

First, the identity 
$\nabla_\mu F^{\mu\nu}=\partial_\mu(\sqrt{-g}F^{\mu\nu})/\sqrt{-g}$ 
leads to the wave equation for the polarization $A_z(t,x,y)$ 
\bea
\label{A_z}
\left(
\partial_t^2
-\partial_x[1-h]\partial_x
-\partial_y[1+h]\partial_y
\right)A_z=0
\,.
\ea
After a spatial Fourier transformation, this reduces to the differential 
equation $\ddot A_z+\Omega^2A_z=0$ 
of a parametric harmonic oscillator with the 
time-dependent frequency~$\Omega(t)$ given by Eq.~\eqref{dispersion} for 
$K_z=0$. 
Since the frequency $\Omega$ of the electromagnetic waves 
(e.g., optical lasers) is much larger than that of the gravitational waves 
$\omega\ll\Omega$, we may employ the WKB approximation and deduce a 
scaling of the amplitude $A_z$ with $1/\sqrt{\Omega}$.
One way to obtain this result is to consider the conserved Wronskian which 
reads $W=A_z^*\dot A_z-\dot A_z^* A_z$ and thus simplifies to 
$W\approx-2i\Omega|A_z^2|$. 

Second, let us consider the fixed polarization $A_x(t,y,z)$, 
for which we find the wave equation 
\bea
\label{A_x}
\left(
\partial_t[1-h]\partial_t
-\partial_y^2
-\partial_z[1-h]\partial_z
\right)A_x=0
\,.
\ea
In this case, the conserved Wronskian contains an additional metric factor 
$W=[1-h](A_x^*\dot A_x-\dot A_x^* A_x)$ and thus the amplitude $A_x$ 
scales with $1/\sqrt{[1-h]\Omega}$.  

Obviously, the third case $A_y(t,x,z)$ is completely analogous to the second 
after replacing $1-h$ by $1+h$.
The behavior of general polarizations $A_i$ can be inferred from the 
wave equation in temporal gauge $A_0=0$ 
\bea
\partial_t\left(g^{ij}\partial_tA_j\right)=K_2^{ij}A_j
\,,
\ea
where the matrix $K_2^{ij}$ contains bilinear forms of the wave-numbers 
$K_i$ as well as metric factors $1\pm h$. 
In this case, the conserved Wronskian reads 
$W=A^*_i g^{ij} \dot A_j - \dot A^*_i g^{ij} A_j$
which can again be used to infer the scaling of the amplitude $A_i$.
Note, however, that the transversality condition $K_i g^{ij} \dot A_j=0$ 
implies small changes of the polarization direction induced by the 
gravitational wave -- unless $K_i$ or $A_j$ are oriented along the 
eigenvectors of $g_{ij}$, i.e., the coordinate axes. 

In summary, both the frequency $\Omega$ as well as the amplitudes $A_i$ 
of the electromagnetic waves acquire small corrections of the form 
$1+\zeta h$ due to the gravitational wave, where the various values of 
$\zeta$ depend on the propagation and polarization directions of the 
electromagnetic waves. 

\section{Matter Hamiltonian}\label{Matter Hamiltonian}

In flat space-time, i.e., without the gravitational wave, we assume that the 
matter 
can be described by the standard non-relativistic many-body Hamiltonian 
\bea
\label{H_0}
\hat H_0
&=&
\int \mathrm{d}^3r 
\left[\frac{1}{2m}(\na\hat\Psi^\dagger)\cdot(\na\hat\Psi)+V_0(\f{r}) 
\hat\Psi^\dagger\hat\Psi
\right] 
\nn
&&+
\frac12
\int \mathrm{d}^3r\,\mathrm{d}^3r'\, 
\hat\Psi^\dagger(\f{r})\hat\Psi^\dagger(\f{r}') 
W(\f{r},\f{r}')
\hat\Psi(\f{r}')\hat\Psi(\f{r}) 
\,,\quad\quad
\ea
with bosonic or fermionic field operators $\hat\Psi$ and $\hat\Psi^\dagger$, 
the static trapping potential $V_0$ and the interaction $W$. 

In order to describe the response 
to a gravitational wave, 
we first have to determine the corresponding changes in the Hamiltonian. 
As already shown in \cite{Schutzhold-BEC,Visser}, for example, the kinetic term is modified quite 
intuitively by inserting the metric $g^{ij}$ into the scalar product between 
the field gradients, i.e., $(\na\hat\Psi^\dagger)\cdot(\na\hat\Psi)$
is replaced by 
$-(\partial_i\hat\Psi^\dagger)g^{ij}(\partial_j\hat\Psi)$. 
The change of the trapping potential $V$ will be discussed below. 
Assuming that the interaction $W$ is isotropic and short-ranged, 
we neglect its modification due to the gravitational wave. 

\subsection{Energy Transfer}\label{Energy Transfer} 

Now we are in the position to study how the energy of the matter 
changes due to its interaction with the gravitational wave. 
To this end, we employ the Heisenberg picture where 
\bea
\label{Heisenberg}
\frac{\mathrm{d}\hat{H}}{\mathrm{d}t}
&=&
\left(\frac{\partial\hat H}{\partial t}\right)_{\rm expl}
=
\frac{\partial\hat H}{\partial h}\,\dot h
\\
&=&
\dot h
\int \mathrm{d}^3r 
\left[
\frac{\partial V}{\partial h}
\hat\Psi^\dagger\hat\Psi+
\frac{(\partial_y\hat\Psi^\dagger)(\partial_y\hat\Psi)-
(\partial_x\hat\Psi^\dagger)(\partial_x\hat\Psi)}{2m}
\right]
\,.
\nonumber
\ea
Taking expectation values yields the change of the total energy 
$E=\langle\hat H\rangle$. 
The very general expression~\eqref{Heisenberg} already allows us 
to infer important consequences. 
In analogy to time-dependent perturbation theory, we may replace 
the expectation values 
(such as $\langle\hat\Psi^\dagger\hat\Psi\rangle$)
in the above integrand to lowest order in $h$
by their undisturbed expressions 
(such as $\langle\hat\Psi^\dagger\hat\Psi\rangle_0$) in flat space-time
because there is already a factor of $\dot h$ in front of the integral 
\bea
\label{undisturbed}
\dot E
&=&
\dot h
\int \mathrm{d}^3r 
\left\langle
\frac{\partial V}{\partial h}
\hat\Psi^\dagger\hat\Psi
+
\frac{(\partial_y\hat\Psi^\dagger)(\partial_y\hat\Psi)
-
(\partial_x\hat\Psi^\dagger)(\partial_x\hat\Psi)}{2m}
\right\rangle_0 
\nn
&&
+\ord(h^2)
\,.
\ea
As a result, if this undisturbed (i.e., initial) state is a stationary state 
with respect to the $\hat H_0$-dynamics --
such as the ground state or a thermal equilibrium state -- 
the above expectation value would be independent of time.
In this case, the time integration of Eq.~\eqref{undisturbed} becomes trivial 
and thus there is no energy shift to linear order in $h$. 
In order to obtain such a first-order energy shift, one should prepare 
a non-stationary state 
(e.g., vibrating or oscillating)
such that the expectation values oscillate -- 
ideally in resonance with $\dot h$ to maximize the energy transfer. 

As another consequence of the general expression~\eqref{Heisenberg}, we may 
estimate the maximum amount of energy which can be transfered. 
To this end, we exploit the non-negativity of the operators 
$\hat\Psi^\dagger\hat\Psi$ and 
$(\partial_i\hat\Psi^\dagger)(\partial_i\hat\Psi)$
which allows us to derive the rigorous upper bound 
\bea
\label{rigorous}
\dot E\leq|\dot h|_{\rm max}  
\left(
\left|\frac{\partial V}{\partial h}\right|_{\rm max}
\langle\hat N\rangle+
\langle\hat E_{\rm kin}\rangle_{\rm max}
\right) 
\,,
\ea
in terms of the total particle number $\langle\hat N\rangle$ and the 
kinetic energy $\langle\hat E_{\rm kin}\rangle$ of the matter. 
Note that the former is conserved, i.e., $\langle\hat N\rangle$ is constant, 
while the latter $\langle\hat E_{\rm kin}\rangle$ may vary with time due to 
an exchange between kinetic, potential and interaction energy.

\subsection{Electromagnetic Analogy}\label{Electromagnetic Analogy} 

It might be illuminating to compare the energy transfer by gravitational waves 
discussed above to the well-known case of electromagnetic waves. 
Again assuming that our quantum system is much smaller than the wavelength
of the electromagnetic field (dipole approximation), we may effectively 
describe it by a purely time-dependent vector potential $\f{A}(t)$.
Then the Hamiltonian~\eqref{H_0} becomes 
\bea
\label{H_0-A}
\hat H_0
&=&
\frac{1}{2m}
\int \mathrm{d}^3r 
\left([\na+iq\f{A}]\hat\Psi^\dagger\right)
\cdot
\left([\na-iq\f{A}]\hat\Psi\right)
\nn
&&+
\int \mathrm{d}^3r\,
V_0(\f{r}) 
\hat\Psi^\dagger\hat\Psi
\nn
&&+
\frac12
\int \mathrm{d}^3r\,\mathrm{d}^3r'\, 
\hat\Psi^\dagger(\f{r})\hat\Psi^\dagger(\f{r}') 
W(\f{r},\f{r}')
\hat\Psi(\f{r}')\hat\Psi(\f{r}) 
\,.\quad\quad
\ea
If we assume that the electromagnetic field does neither affect the 
potential $V_0(\f{r})$ nor the interaction $W(\f{r},\f{r}')$, 
the analogue of Eq.~\eqref{Heisenberg} reads 
\bea
\label{poynting}
\dot E=\dot{\f{A}\;}\cdot\int \mathrm{d}^3r
\left[
iq\,\frac{\hat\Psi^\dagger\na\hat\Psi-(\na\hat\Psi^\dagger)\hat\Psi}{2m}
+q^2\frac{\f{A}\hat\Psi^\dagger\hat\Psi}{m}\,
\right]
.\quad\quad 
\ea
For a purely time-dependent vector potential $\f{A}(t)$, the second term 
$\propto q^2$ yields the total particle number $\hat N$. 
Since $\hat N$ is conserved, this term does not generate a net energy shift.
The same line of reasoning would apply to the term $\partial V/\partial h$ 
in Eq.~\eqref{Heisenberg} if $\partial V/\partial h$ was purely time-dependent. 
Still, it is advantageous to keep this second term $\propto q^2$ in order to 
retain gauge invariance. 

Altogether, we find that Eq.~\eqref{Heisenberg} is analogous to the well-known 
Poynting theorem in electrodynamics as the integrand of Eq.~\eqref{poynting}
represents the current density $\f{j}$. 
Thus, $\dot E$ can be bound in analogy to Eq.~\eqref{rigorous} by 
electric field $|\dot{\f{A}}|_{\rm max}$, 
current density $|\f{j}|_{\rm max}$, and volume.  

\section{Toy Model}\label{Toy Model} 

In order to understand the above result~\eqref{Heisenberg} 
by means of a simple toy model,
let us consider two classical and non-relativistic point particles of mass $m$
on circular orbits around their joint center of mass 
\bea 
\f{r}_\pm(t)
=
\pm R
\left(
\begin{array}{c}
\cos(\omega_{\rm rot}t) \\ 
\sin(\omega_{\rm rot}t) \\ 
0
\end{array}
\right) 
\,.
\ea
Besides the force holding the masses on their circular orbits, 
the gravitational wave induces a small additional acceleration, 
as given by the geodesic equations already discussed in 
Sec.~\ref{Massive Particles}, i.e., 
$\dot u^x=-\dot h u^x$ and $\dot u^y=\dot h u^y$ as well as 
$\dot u^t=\dot h(u_y^2-u_x^2)/2$.  
The resulting change in energy is thus given by 
$\dot E=\dot h(E_{\rm kin}^y-E_{\rm kin}^x)$, 
in analogy to Eq.~\eqref{undisturbed}. 
If the frequency $\omega$ of the gravitational wave equals twice the rotational
frequency $\omega_{\rm rot}$, we obtain a resonant transfer of energy, 
see Appendix~\ref{Rotating Frame} and 
\cite{Braginskii-1967,Braginskii-1969} as well as \cite{Press} 
and references therein.

It might be illuminating to insert some numbers and to estimate the resulting 
orders of magnitude.
Assuming a gravitational wave with a frequency $\omega$ in the kHz regime and an amplitude of $h=\ord(10^{-22})$, we may estimate the energy $\Delta E$ 
transferred after an interaction time $T$ of one hundred cycles, i.e., 
$\omega T=\ord(10^2)$.
Then, demanding that this energy shift $\Delta E=\ord(h\omega T E_{\rm kin})$ 
corresponds to one excitation quantum $\hbar\omega$ in the kHz regime, 
we would need an initial kinetic energy of order $10^8~\rm eV$ or 
$10^{-11}~\rm J$. 

Even though it would be easy to prepare such an initial kinetic energy for 
mesoscopic or macroscopic matter distributions, actually detecting an energy 
shift of one excitation quantum $\hbar\omega$ on top of this huge background 
is certainly extremely challenging.
As a way around this obstacle, one could consider the change in vibrational 
energy $E_{\rm vib}$ instead of rotational energy $E_{\rm rot}$.
The acceleration induced by the gravitational wave has also components in 
radial direction, which lead to a change in vibrational energy of order 
\bea
\label{Delta-Evib}
\Delta E_{\rm vib}=\ord\left(h\omega T\sqrt{E_{\rm vib}E_{\rm rot}}\right)
\,,
\ea
if an initial vibration of the barbell is present, i.e., $\dot{R}\neq 0$. 
In order to obtain resonant energy transfer, the vibrational frequency
$\omega_{\rm vib}$ should match $|\omega\pm2\omega_{\rm rot}|$. 
In the following, we assume that all three frequencies are in the kHz regime. 
Then, if the initial quantum state of the vibrational mode corresponds to a few 
excitation quanta (say, ten $\hbar\omega_{\rm vib}$), an energy shift of 
one excitation quantum $\hbar\omega_{\rm vib}$ would require a rotational energy $E_{\rm rot}=10^8~\rm J$. 

Of course, this value is now much larger than in the previous case 
($10^8~\rm eV$ or $10^{-11}~\rm J$), but it is not completely out of reach. 
For example, a barbell with $m=\ord(100~\rm kg)$ and $R=\ord(\rm m)$, 
rotating with $\omega_{\rm rot}=\ord(\rm kHz)$, would have such a 
rotational energy $E_{\rm rot}=\ord(10^8~\rm J)$. 
Obviously, controlling the vibrational modes to the desired accuracy would 
still be very challenging and probably requires a barbell levitating or 
suspended in ultra-high vacuum etc. 
On the other hand, the impressive experimental progress regarding 
controlling and cooling down vibrational modes of macroscopic objects 
(see, e.g., \cite{Whittle,Zoepfl}) gives rise to the hope that such an experiment may 
not be totally out of reach. 

As an alternative, one could envision two concentric and co-rotating 
barbells at a right angle and consider the scissors-like motion instead
of the vibrational mode.
In doing so, one can find basically the same energy transfer as given by Eq.~\eqref{Delta-Evib}.

\section{Bose-Einstein Condensates} 

\subsection{Trapping Potential} 

After this simple toy model, 
let us apply our results to Bose-Einstein condensates. 
To this end, we first have to determine how the trapping potential $V$ changes. 
As already mentioned, this will depend on its explicit physical realization 
in general. 
As an extreme case, if the shape of $V$ is only determined by the positions
of effectively force-free masses at rest (such as the mirrors in LIGO), 
it would not change at all during the passage of a gravitational wave. 

However, for more realistic scenarios, one would expect $V$ to vary. 
As a concrete example, let us consider optical traps which are often used 
to confine Bose-Einstein condensates. 
They may consist of a superposition of standing laser beams in various 
directions.  
As discussed in Sec.~\ref{Electromagnetic Waves}, these electromagnetic 
waves respond to gravitational waves via modification factors of the form 
$1+\zeta h$ 
in front of their frequencies and amplitudes where the $\zeta$ values 
are typically of order unity and depend on polarization and propagation 
direction.

The atoms in the Bose-Einstein condensate
are then polarized by the electromagnetic waves 
where their polarizability scales with $1/(\Omega^2-\Omega_{\rm res}^2)$
in terms of the frequencies $\Omega$ of the electromagnetic wave and 
the relevant atomic resonance $\Omega_{\rm res}$
(blue or red detuned atoms).
Assuming that the $\Omega_{\rm res}$ do not change (see Appendix~\ref{Atomic Eigenstates}),
these polarizabilities get also modified by the gravitational wave via 
the change in $\Omega$. 
Actually, if $\Omega$ is close to the resonance frequency $\Omega_{\rm res}$, 
the response to gravitational waves is enhanced, but going too close to 
resonance can be problematic. 

In addition to these effects already occurring for free electromagnetic 
waves, one should also include their sources and boundary conditions 
(i.e., mirrors) which may induce further factors of $1+\zeta h$.
Since these various factors of $1+\zeta h$ stem from different effects,
their values of $\zeta$ will typically be different and hence they will 
not cancel each other in general. 

In order to accommodate all these different factors of $1+\zeta h$,
we employ the standard harmonic approximation 
$V_0(\f{r})=\f{r}\cdot\f{M}_0\cdot\f{r}$
for the trapping potential $V_0$ near its minimum 
(which we set to $\f{r}=0$) with some matrix $\f{M}_0$.
Then, in view of the above considerations, 
the most general form for the modifications due to the gravitational
wave can be cast into the form 
\bea
\label{potential-modifications}
V(t,\f{r})
=
\f{r}\cdot(\f{M}_0+h\f{M}_1)\cdot\f{r}
+
h\f{F}_1\cdot\f{r}
+hV_1 
\,.
\ea
The perturbations $\f{M}_1$, $\f{F}_1$ and $V_1$ account for all 
the factors $1+\zeta h$ mentioned above and thus depend on the 
amplitudes, polarizations, frequencies and propagation directions 
of the various laser beams as well as the atomic resonances 
(and the mirrors etc.). 

In addition to the modification $\f{M}_1$ of the shape of the potential,
which one would naturally expect from a gravitational wave, one can also 
have a shift in position $\f{F}_1$ (in asymmetric scenarios) and in energy 
$V_1$.
Since the total particle number $\hat N$ commutes with the Hamiltonian, 
the term $V_1$ has no effect (unless we observe interference between 
two Bose-Einstein condensates with different $V_1$).

\subsection{Excitations} 

In order to study the excitations in the Bose-Einstein condensate induced by the 
gravitational wave, we employ the standard mean-field approximation 
$\hat\Psi\to\psi_{\rm c}+\delta\psi$ where 
$\psi_{\rm c}$ denotes the undisturbed wave-function of the condensate 
(i.e., in the absence of the gravitational wave) while $\delta\psi$ are the 
perturbations.
Linearizing in $\delta\psi$ then yields the Bogoliubov-de~Gennes equations 
which now acquire a source term due to the gravitational wave 
\bea 
\label{BdG}
\left(
i\partial_t
+ \frac{\na^2}{2m} 
-V_0 
-2g |\psi_{\rm c}|^2 
\right)\delta\psi 
-g\psi_{\rm c}^2 \delta\psi^*
= 
\nn
h\left(
\frac{\partial_y^2-\partial_x^2}{2m}
+\frac{\partial V}{\partial h}
\right)\psi_{\rm c}
\,.\quad 
\ea 
Assuming rotational symmetry for the undisturbed condensate 
(i.e., for $\psi_{\rm c}$ and $V_0$), we find that the direct interaction  
$\propto(\partial_y^2 - \partial_x^2)$ in the first term of the second line 
in Eq.~\eqref{BdG} generates quadrupolar excitations $\delta\psi$, 
as expected from a gravitational wave. 
However, the indirect interaction via changes in the trapping potential 
may also generate other (e.g., dipolar) excitations $\delta\psi$, 
provided that such contributions (e.g., $\f{F}_1$) occur in 
Eq.~\eqref{potential-modifications}.

In order to make the connection to fluid dynamics more apparent, 
we use the Madelung split $\psi_{\rm c}=\sqrt{\rho}\,e^{iS}$ in 
terms of condensate density $\rho$ and phase $S$ where the 
perturbation $\delta\psi$ is then represented by $\delta\rho$ 
and $\delta S$.
In this form, Eq.~\eqref{BdG} splits into two real equations 
\begin{equation}
\label{eq:continuity-GW-lin}
(\partial_t + \na\cdot\f{v})\delta\rho 
+\na\cdot\left(\frac{\rho}{m}\na\delta S\right) 
= 
h\left[\partial_y(\rho v_y) - \partial_x(\rho v_x)\right]
\end{equation}
and (re-inserting $\hbar$ for the discussion below)
\begin{align}
\label{eq:Hamilton-Jacobi-GW-lin}
(\partial_t + \f{v}\cdot\na)\delta S 
+ g \delta \rho  + 
\frac{\hbar^2}{4m}
\frac{\delta\rho\na^2\sqrt{\rho}-\rho\na^2(\delta\rho/\sqrt{\rho})}
{\rho^{3/2}}
=
\nn
h\left[
\frac{m}{2}\left(v_y^2-v_x^2\right)-
\frac{\partial V}{\partial h}+
\frac{\hbar^2}{2m}
\frac{\partial_x^2\sqrt{\rho}-\partial_y^2\sqrt{\rho}}{\sqrt{\rho}}\right],
\end{align}
where $\f{v}=\na S/m$ is the condensate velocity. 

For length scales much larger than the healing length, we may neglect the 
``quantum-pressure'' terms $\propto\hbar^2$ in 
Eq.~\eqref{eq:Hamilton-Jacobi-GW-lin} 
such that the two first-order equations above can be combined into one 
second-order equation
\begin{align}
\label{eq:wave-equation-GW}
(\partial_t+\na\cdot\f{v})(\partial_t+\f{v}\cdot\na)\delta S 
- 
\na\cdot\left(\frac{g\rho}{m}\na\delta S\right) 
=
\nn
\left[
\frac{m}{2}(\partial_t+\na\cdot\f{v})\left(v_y^2-v_x^2\right)
-(\partial_t+\na\cdot\f{v})\frac{\partial V}{\partial h}
\right]h
\nn
+gh\left[\partial_x(\rho v_x)-\partial_y(\rho v_y)\right] 
\,.
\end{align}
As an extremely simple example, we may consider homogeneous condensates 
at rest for which the above equation simplifies to 
$(\partial_t^2-c_{\rm s}^2\na^2)\delta S=\dot h\partial V/\partial h$
with the speed of sound $c_{\rm s}^2=g\rho/m$.
In this case, the generated fluctuations $\delta S=\ord(h)$ can be obtained 
via the well-known retarded Green function of the d'Alembertian. 
Note, however, that these first-order fluctuations $\delta S=\ord(h)$
do not generate a first-order energy shift because the background state 
is stationary, as explained in Sec.~\ref{Energy Transfer}. 

\subsection{Estimate of Energy Transfer} 

Finally, let us exemplify the rigorous bound~\eqref{rigorous} for a general 
non-stationary state of a Bose-Einstein condensate. 
As in Sec.~\ref{Toy Model},  
the first factor $\dot h$ can be estimated by the typical frequencies 
$\omega=\ord(\rm kHz)$ and amplitudes $h=\ord(10^{-22})$ of gravitational waves. 

In order to estimate the derivative $\partial V/\partial h$, we may start from 
the harmonic approximation~\eqref{potential-modifications}.
For optical traps, the order of magnitude of the potential strength is set 
by the recoil energy $E_{\rm R}=k^2/(2m)$ which is typically in the $\mu\rm K$ 
regime.   
Here $k=2\pi/\lambda$ is the momentum of the photons forming the optical 
trap and $m$ the mass of the trapped atoms (e.g., rubidium).  
In the absence of further large numbers, one would expect that 
$\f{M}_0$ and $\f{M}_1$ in Eq.~\eqref{potential-modifications}
scale with $\ord(E_{\rm R}/\lambda^2)$ 
while $\f{F}_1=\ord(E_{\rm R}/\lambda)$ and $V_1=\ord(E_{\rm R})$.
Of course, extending the harmonic approximation~\eqref{potential-modifications}
to large distances $\f{r}$, the derivative $\partial V/\partial h$ would grow
formally without any bound. 
This artifact can be avoided by limiting the maximum distance $\f{r}$ to the 
size of the condensate or the region of applicability of the 
harmonic approximation~\eqref{potential-modifications}.
Both are set by the optical wavelength $\lambda=\ord(\mu\rm m)$ such that we 
arrive at $\partial V/\partial h=\ord(E_{\rm R})$.

The remaining term in Eq.~\eqref{rigorous} is the kinetic energy 
$\langle\hat E_{\rm kin}\rangle_{\rm max}$ of the condensate. 
Obviously, the maximum kinetic energy per atom should not exceed the total
potential depth of order $\ord(E_{\rm R})$ in order to stay trapped. 
Thus we have $\langle\hat E_{\rm kin}\rangle_{\rm max}\leq\ord(NE_{\rm R})$
where $N=\langle\hat N\rangle$ is the total number of atoms in the condensate.

Altogether we arrive at the following order-of-magnitude estimate for the 
energy shift 
\bea
\label{order-of-magnitude}
\Delta E\leq\ord\left(h\omega TNE_{\rm R} \right) 
\,,
\ea
where $T$ is the interaction time 
(e.g., the duration of the gravitational wave). 
Inserting a typical amplitude $h=\ord(10^{-22})$, 
a number of cycles $\omega T=\ord(10^2)$, 
a rather large atom number $N=\ord(10^9)$, 
and a characteristic potential strength $E_{\rm R}=\ord(\mu\rm K)$, 
we find an energy shift $\Delta E$ in the atto-Kelvin regime -- 
which is probably too small to be measurable. 
Note that this is not the energy shift per particle, but the energy 
shift for the whole condensate. 

Turning this argument~\eqref{order-of-magnitude} around, 
an  energy shift of $\Delta E=\ord(10~\rm nK)$ 
corresponding to the energy $\hbar\omega$
of a single kHz-phonon would require a 
characteristic potential strength (and energy per atom) of order 
ten Kelvin, which is also beyond current experimental capabilities.

\section{Conclusions} 

We study the interaction between gravitational waves and quantum matter 
and find two major coupling mechanisms. 
First, the gravitational wave encoded in the metric $g^{ij}$
directly affects the kinetic term $(\na\hat\Psi^\dagger)\cdot(\na\hat\Psi)$
which is replaced by $-(\partial_i\hat\Psi^\dagger)g^{ij}(\partial_j\hat\Psi)$. 
Second, the gravitational wave may indirectly couple to matter 
by modifying its trapping potential $V$ (see Appendix~\ref{Atomic Eigenstates}).

As a possible observable, we consider the energy transfer $\Delta E$ between 
the gravitational wave and matter. 
For stationary initial states, 
we find that this energy transfer 
$\Delta E$ vanishes to first order in the amplitude $h$ of the gravitational wave.
For arbitrary initial states, 
we derive a general rigorous bound for the energy transfer $\Delta E$ 
in terms of particle number and initial kinetic energy. 

As a first example, we discuss a simple toy model in the form of a rotating 
barbell. 
For quite moderate rotational energies $E_{\rm rot}$, the energy transfer 
$\Delta E_{\rm rot}$ 
can exceed one excitation quantum $\hbar\omega$, but actually measuring 
this small change on top of a huge background $E_{\rm rot}$ is very 
challenging.
As a possible remedy, one might consider the change of the vibrational 
energy $\Delta E_{\rm vib}$ instead. 
Demanding that this change $E_{\rm vib}$ exceeds one excitation quantum 
$\hbar\omega$ then requires a rotational energy $E_{\rm rot}$ which is 
much larger 
(assuming a reasonably small initial vibrational energy $E_{\rm vib}$),
but not necessarily out of reach. 

As a second example, we apply our findings to Bose-Einstein condensates, 
where we discuss the gravitationally induced modifications of the trapping 
potential $V$ for the explicit example of optical traps. 
Assuming rotational symmetry of the undisturbed condensate, we find that the 
direct interaction mechanism involving the kinetic term generates quadrupolar 
excitations (as expected) while the indirect coupling via the potential $V$
may also induce other (e.g., dipolar) excitations -- depending on the specific 
realization of the trap. 

Quite generally, inserting typical orders of magnitude of Bose-Einstein 
condensates into the rigorous bound for the energy transfer $\Delta E$, 
we find that it is probably far too small to be detectable with present-day 
technology -- at least in absence of further large numbers which may enhance 
the signal.  

\section{Outlook} 

As we may infer from the rigorous bound, one way to increase the possible
energy transfer $\Delta E$ could be to consider other forms of matter 
such as super-fluid helium or ultra-cold solids containing more particles and 
thus admitting higher kinetic energies. 
For example, one could envisage levitating helium droplets or barbells 
which display quadrupolar vibrations or rotations in resonance with the 
gravitational wave. 
In this case, it might be easier to achieve an energy transfer $\Delta E$ 
corresponding to one or more excitation quanta $\hbar\omega$. 
Of course, detecting such a small change of energy experimentally is another 
challenge. 

Going a bit further, let us discuss these scenarios in some more detail.
If the levitating helium droplets or barbells display quadrupolar vibrations 
or rotations in resonance with the gravitational wave, the sign of their energy 
shift $\Delta E$ depends on the relative phase between the gravitational wave 
and the quadrupolar vibration or rotation. 
If they are in phase, the energy increases $\Delta E>0$ 
but if they are out of phase (by a phase shift of $\Delta\varphi=\pi$), 
the energy decreases $\Delta E<0$.
In analogy to photons as quanta of electromagnetic waves, we may use the 
picture of gravitons as quanta of gravitational waves.
Then, the first case corresponds to the absorption of gravitons, 
while the second scenario describes the stimulated emission of gravitons.
Such a stimulated emission scenario 
may be our best chance to actually emit 
gravitons in a controlled earth-bound experiment -- but it would still 
be a challenging experiment. 
However, it would mark the important step from merely observing 
a natural phenomenon to actually manipulating it. 
Of course, detecting the gravitons emitted in this way would then be yet 
another challenge. 

\acknowledgments 

The authors thank F.~Queisser and W.G.~Unruh
for fruitful discussions and acknowledge funding 
by the Deutsche Forschungsgemeinschaft (DFG, German Research Foundation) 
-- Project-ID 278162697-- SFB 1242. 

\appendix 

\section{Atomic Eigenstates}\label{Atomic Eigenstates} 

For the sake of completeness and as another illustration for the impact of 
gravitational waves, let us investigate the induced modifications of the 
atomic eigenstates. 
For simplicity, let us start with the non-relativistic hydrogen atom as 
described by the undisturbed Hamiltonian 
\bea
\label{app-H0}
\hat H_0=\frac{\hat{\f{p}}^2}{2m}+V(\hat r)
\,,
\ea
where $V(\hat r)=-q^2/(4\pi\hat r)$ denotes the Coulomb potential. 
Then, in complete analogy to the Hamiltonian~\eqref{H_0}, 
the impact of the gravitational wave can be encoded in the perturbation
Hamiltonian \cite{Wicht,footnote}
\bea
\label{app-H1}
\hat H_1=
h\left[
\frac{\hat p_y^2-\hat p_x^2}{2m}
+
q^2\frac{\hat x^2-\hat y^2}{8\pi\hat r^3}
\right] 
=\hat H_1^{\rm kin}+\hat H_1^{\rm pot}.
\;
\ea
The deformation of the Coulomb potential can be derived via replacing the 
flat space-time Laplace operator $\na^2$ in the Poisson equation for 
$V(\f{r})$ by the Laplace-Beltrami operator $-\partial_i g^{ij} \partial_j$, see also \cite{Quinones}.  

Since $h$ is slowly varying in comparison to the atomic frequencies, 
we may estimate the lowest-order variations of the eigenstates via 
stationary perturbation theory. 
The first-order shift of the eigenenergies is determined by the 
expectation values of the perturbation Hamiltonian~\eqref{app-H1} 
in the undisturbed eigenstates. 
Obviously, the expectation value in the 1s ground state vanishes in view 
of rotational invariance $\bra{{\rm 1s}}\hat H_1\ket{{\rm 1s}}=0$.
More generally, matrix elements $\bra{n,\ell,m}\hat H_1\ket{n',\ell',m'}$
can only yield non-vanishing contributions if $m'=m\pm2$.
Thus, one might expect an energy shift for ${\rm p}_x$-orbitals, for example, 
see also \cite{Wanwieng}.
Indeed, the expectation value of $\hat H_1^{\rm kin}$ corresponds to the 
difference between the average kinetic energies in $x$- and $y$-direction 
and yields a non-zero result 
\bea
\label{app-kinetic}
\bra{{\rm 2p}_x}\hat H_1^{\rm kin}\ket{{\rm 2p}_x}
=-h\,\frac{q^2}{80\pi a_{\rm B}}
\,,
\ea
where $a_{\rm B}$ is the Bohr radius. 
Apart from the small pre-factor $h$, this energy shift is in the eV regime 
and thus one might expect it to be measurable. 
However, one should not forget the second contribution $\hat H_1^{\rm pot}$.
Calculating its expectation value 
$\bra{{\rm 2p}_x}\hat H_1^{\rm pot}\ket{{\rm 2p}_x}$ one finds that it 
precisely cancels the above contribution~\eqref{app-kinetic}  
leading to a vanishing energy shift 
$\bra{{\rm 2p}_x}\hat H_1\ket{{\rm 2p}_x}=0$
to lowest order, consistent with the results of \cite{Parker-1980,Parker-1980-D,Parker-1982,Pinto-1993}.

This cancellation is perhaps not too surprising because a constant $h$ 
can be interpreted as a trivial change of coordinates $x\to[1+h/2]x$ 
and $y\to[1-h/2]y$, which should not affect any physical quantities 
such as energies.
However, such a change of coordinates is consistent with 
modifications of the wave-functions and thus non-diagonal matrix elements 
can be non-vanishing, such as 
\bea
\bra{{\rm 1s}}\hat H_1\ket{{\rm 3d}_{x^2-y^2}}=
h\,\frac{q^2}{128\pi a_{\rm B}}
\,,
\ea
see also \cite{Fischer,Pinto}. As a consequence, transition matrix elements could also change 
(in those coordinates).

In view of the above argument based on coordinate independence 
(i.e., general covariance), the cancellation of the energy shifts to lowest 
order in $h$ should remain valid in the general case.
As an example, let us briefly discuss the Dirac equation. 
To lowest order in $h$, the metric~\eqref{metric} can be incorporated by a 
modification of the Dirac matrices 
$\gamma^x\to[1-h/2]\gamma^x$ and $\gamma^y\to[1+h/2]\gamma^y$
while $\gamma^z$ and $\gamma^t$ remain unchanged. 
Again using that $h$ is slowly varying, we may neglect the Fock-Ivanenko 
(spin connection) coefficients because they scale with the derivative 
$\dot h=\ord(\omega h)$ and are thus suppressed for small~$\omega$. 
As a consequence, the Dirac perturbation Hamiltonian has a structure very 
similar to the Schr\"odinger case~\eqref{app-H1}.
The potential part $\hat H_1^{\rm pot}$ stemming from the deformation of 
the Coulomb potential is basically the same, while the kinetic part 
$\hat H_1^{\rm kin}$ reads $h[\alpha^yi\partial_y-\alpha^xi\partial_x]/2$ 
where the $\alpha^i=\gamma^0\gamma^i$ are the velocity matrices in the 
Dirac representation -- in analogy to the Schr\"odinger case~\eqref{app-H1}.

\begin{widetext}

\section{Rotating Frame}\label{Rotating Frame} 

For studying rotating matter distributions such as the barbell, 
it is often useful to transform into the rotating frame. 
Assuming that potential $V_0$ and interaction $W$ are isotropic, 
the Hamiltonian~\eqref{H_0} in the rotating frame reads 
\bea
\label{Hamiltonian-rotating}
\hat H_0^{\rm rot}
&=&
\int \mathrm{d}^3r 
\left[
\frac{1}{2m}
\left(
\na\hat\Psi^\dagger -im[\f{\omega}_{\rm rot}\times\f{r}] \hat\Psi^\dagger
\right) 
\cdot 
\left(
\na\hat\Psi +im[\f{\omega}_{\rm rot}\times\f{r}] \hat\Psi
\right) 
+V_0(\f{r}) \hat\Psi^\dagger\hat\Psi
-\frac{m}{2}\left(\f{\omega}_{\rm rot}\times\f{r}\right)^2 
\hat\Psi^\dagger\hat\Psi
\right]
\nn
&&+
\frac12
\int \mathrm{d}^3r\,\mathrm{d}^3r'\, 
\hat\Psi^\dagger(\f{r})\hat\Psi^\dagger(\f{r}') 
W(\f{r},\f{r}')
\hat\Psi(\f{r}')\hat\Psi(\f{r}) 
\,.
\ea
While the interaction term in the second line does not change 
(due to the assumed isotropy), the first line now contains the kinetic energy 
in the rotating frame $E_{\rm kin}^{\rm rot}$ plus the effective potential 
$V_{\rm eff}(\f{r})=V_0(\f{r})-m(\f{\omega}_{\rm rot}\times\f{r})^2/2$. 
In order to ensure stability, we assume that $V_0$ is stronger than the 
centrifugal potential $m(\f{\omega}_{\rm rot}\times\f{r})^2/2$.

Quite importantly, the effective Hamiltonian~\eqref{Hamiltonian-rotating} 
is independent of time. 
This allows us to prepare an initial state which is stationary or even static 
in the rotating frame.
Furthermore, the well-known analogy to charged particles in a magnetic field 
described by the effective vector potential 
$\f{A}_{\rm eff}\propto\f{\omega}_{\rm rot}\times\f{r}$
enables us to transfer many of the concepts to the rotating case. 
For example, the conserved current contains an additional term from 
$\f{A}_{\rm eff}$
\bea
\hat{\f{j}}_{\rm rot}=\frac{1}{2mi}
\left[\hat\Psi^\dagger\na\hat\Psi-{\rm h.c.}\right]
+\left(\f{\omega}_{\rm rot}\times\f{r}\right)\hat\Psi^\dagger\hat\Psi
\,.
\ea
Now let us study the impact of the gravitational wave~\eqref{metric}.
Assuming a rotation around the $z$-axis for simplicity with an angle 
of $\varphi(t)=\omega_{\rm rot}t$, 
the induced interaction Hamiltonian becomes 
\bea
\hat H_{\rm int}
=
h(t)
\int \mathrm{d}^3r
\left[
\cos(2\omega_{\rm rot}t)
\frac{(\partial_y\hat\Psi^\dagger)(\partial_y\hat\Psi)-
(\partial_x\hat\Psi^\dagger)(\partial_x\hat\Psi)}{2m}  
-
\sin(2\omega_{\rm rot}t)
\frac{(\partial_x\hat\Psi^\dagger)(\partial_y\hat\Psi)+
(\partial_y\hat\Psi^\dagger)(\partial_x\hat\Psi)}{2m}
\right] 
\,,
\ea
where we have again omitted the changes of $V_0$ and $W$ 
induced by the gravitational wave. 

Now let us estimate the energy transfer in analogy to Sec.~\ref{Energy Transfer}.
As an important difference to that section, the terms such as 
$(\partial_y\hat\Psi^\dagger)(\partial_y\hat\Psi)$ can no longer be directly 
bound by the kinetic energy $E_{\rm kin}^{\rm rot}$ which now contains more 
contributions and is given in the first line of Eq.~\eqref{Hamiltonian-rotating}.
In order to place a bound on these additional terms we assume that the initial 
(unperturbed) state is static in the rotating frame which implies 
$\langle\hat{\f{j}}_{\rm rot}\rangle_0=0$. 
Using this assumption and the Cauchy-Schwarz inequality, 
we finally arrive at 
\bea
\dot E\leq 
|\dot h|_{\rm max} 
\left(
2E_{\rm kin}^{\rm rot}+m\int \mathrm{d}^3r\,
\left(\f{\omega}_{\rm rot}\times\f{r}\right)^2 
\langle\hat\Psi^\dagger\hat\Psi\rangle_0
\right) 
+\ord(h^2)
\,.
\ea
Quite intuitively, apart from the kinetic energy within the rotating frame, 
we also obtain a contribution from the rotation itself -- which can be 
bound by the total particle number and the maximum spatial extend of the 
matter distribution (e.g., barbell).

\end{widetext}


\begin{thebibliography}{99}
\bibitem{footnote}
Note that there is a sign difference in comparison to Ref.~\cite{Wanwieng} which 
probably explains the deviation of the final results.
\end{thebibliography}
\end{document}